\documentclass[reprint]{revtex4-1}

\usepackage{graphicx}
\usepackage{dcolumn}

\begin{document}

\title{Probing the BFKL Pomeron with Future ATLAS Forward Detectors}%

\author{Maciej Trzebi{\'n}ski}
\affiliation{IRFU/Service de physique des particules, CEA/Saclay,\\91191 Gif-sur-Yvette cedex, France}
\affiliation{Institute of Nuclear Physics Polish Academy of Sciences\\ul. Radzikowskiego 152, 31-342 Krak\'ow, Poland}
\email{maciej.trzebinski@cern.ch}

\date{\today}

\begin{abstract}
The Jet-Gap-Jet (JGJ) process as a way to test the Balitsky-Fadin-Kuraev-Lipatov Pomeron is introduced. The comparison with the Tevatron data as well as the predictions for the LHC are presented. In the second part, the possibility to measure the Double Pomeron Exchange (DPE) Jet-Gap-Jet process in the ATLAS experiment with additional AFP detectors is described. The ratio of the DPE jet-gap-jet cross section to the DPE jet cross section is presented.
\end{abstract}

\maketitle

\section{Introduction}
In the majority of the events in hadron-hadron collisions an object exchanged between an interacting partons is a quark or a gluon. Such exchange typically results in a multiparticle final state distributed over several units of rapidity. These types of interactions are not the only possibilities. An object exchanged in $t$-channel can be a colour-singlet. Such process is expected to yield a rapidity gap, \textit{i.e.} a space in rapidity devoid of particles. One of the best candidates for such colour singlet is the Balitsky-Fadin-Kuraev-Lipatov (BFKL) Pomeron \cite{Pomeron}.

\section{Jet-Gap-Jet Process}
\begin{figure}[!htbp]
\centering
\includegraphics[width=0.6\columnwidth]{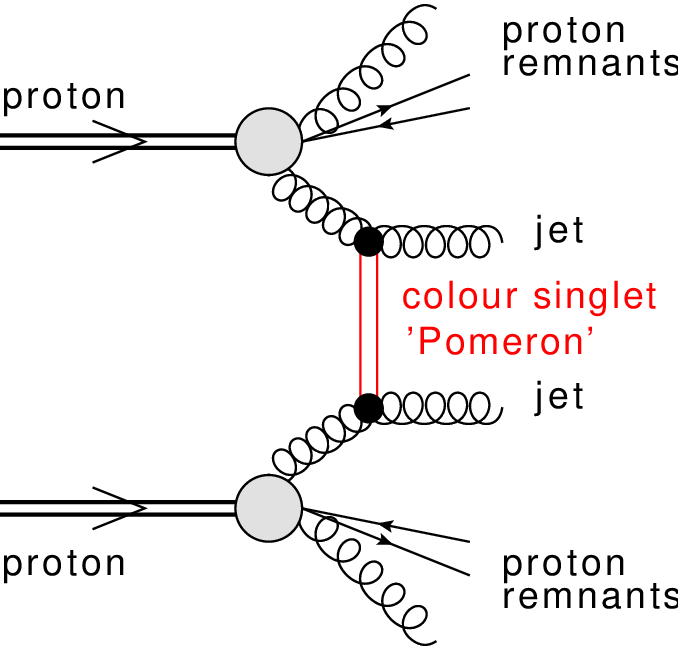}\\[0.2cm]
\includegraphics[width=0.9\columnwidth]{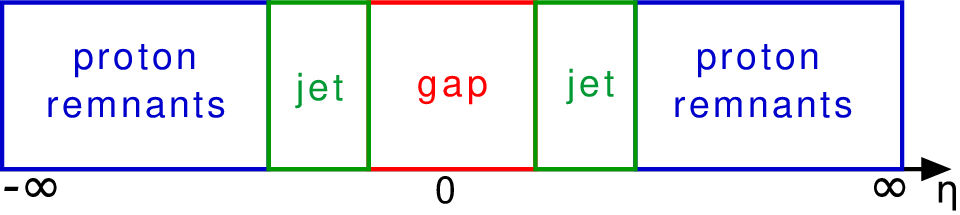}\hfill
\caption{The Jet-Gap-Jet production: the interacting protons are destroyed and two jets are produced. An object exchanged in $t$-channel is a colour singlet, therefore there is a gap in rapidity between two jets.}
\label{fig_diag_JGJ}
\end{figure}

For more than 20 years a large effort has been devoted to understand the QCD dynamics of rapidity gaps in jet events when there is a rapidity gap between two high-$p_{T}$ jets. The diagram of such production is shown in Fig. \ref{fig_diag_JGJ}. One can immediately realise that in such process both protons are destroyed and, due to the colour singlet exchange, no particles are produced between the leading jets.

Despite of the fact that the studies of Jet-Gap-Jet events were performed at the Tevatron, there is still no consensus on what the relevant QCD mechanism really is. The LHC accelerator opens a possibility to shed more light on this topic. The theoretical description of this process is discussed elsewhere (\cite{Royon} and references within), whereas here only the main results are discussed.

\subsection{Comparison with Tevatron Data}
The D0 Collaboration measured the Jet-Gap-Jet event ratio, $R$, defined as the ratio of the Jet-Gap-Jet cross section to the inclusive di-jet cross section:
$$R = \frac{\sigma(JGJ)}{\sigma(Jets)}.$$
This was done as a function of the transverse energy of the second-leading jet, $E_T$, and as a function of the rapidity difference $\Delta\eta_J$ between the two leading jets \cite{D0}. In the analysis, at least two jets were required to be reconstructed in the D0 calorimeter and the $E_T$ of the second leading was greater than 15 GeV. In addition, the leading jets were required to be in the forward regions and in opposite pseudorapidity hemispheres.

The comparison with the BFKL theoretical predictions was performed in \cite{Royon}. The main conclusion was that there is a fair agreement
between the NLL-BFKL calculation and the data whereas the LL-BFKL calculation leads to an $E_T$ dependence which is too flat. However, due to the large experimental errors, the repetition of this test at the LHC accelerator is much desirable.

\begin{figure}[htb]
\centering
\includegraphics[width=1.0\columnwidth]{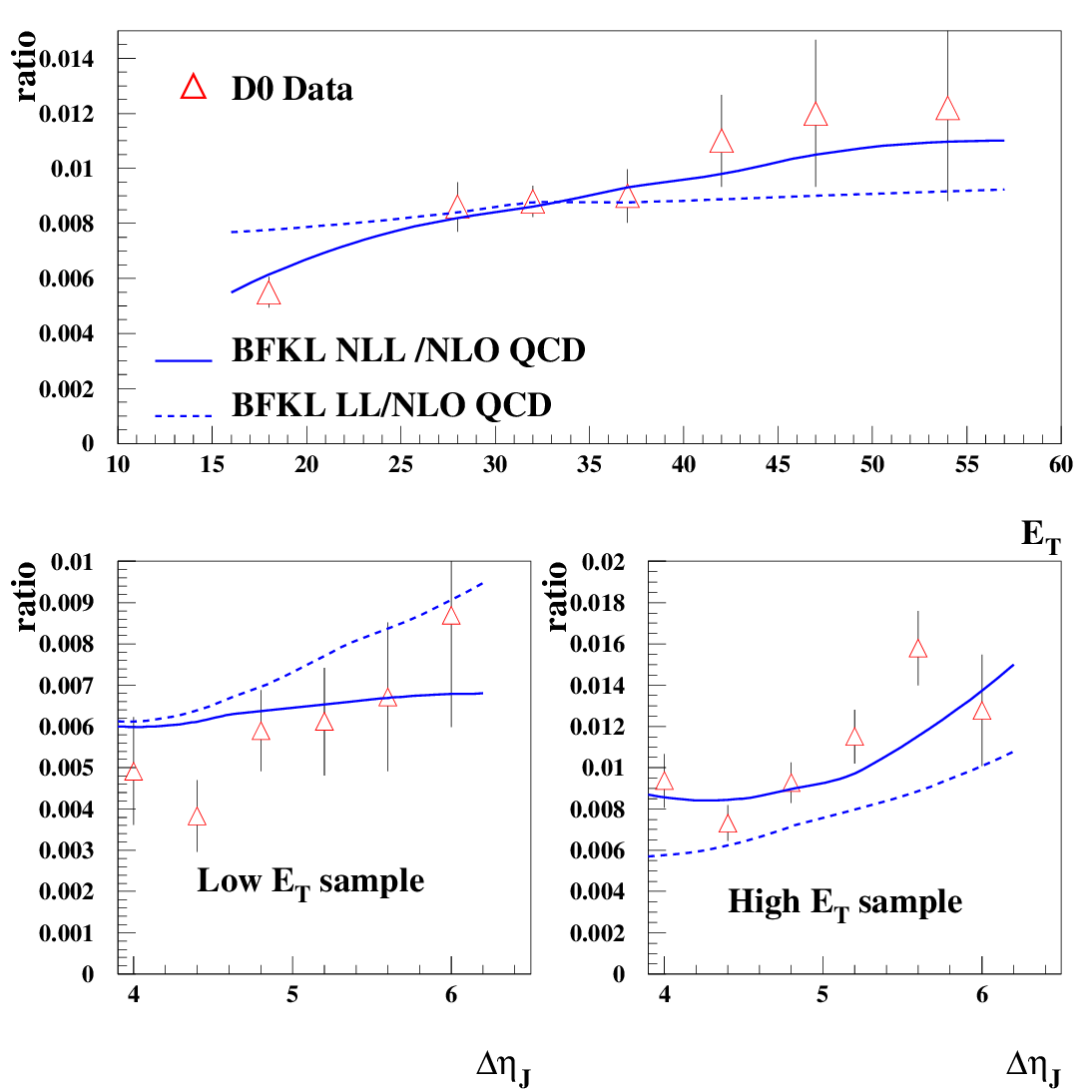}
\caption{Comparisons between the D0 measurements of the jet-gap-jet event ratio with the NLL- and LL-BFKL calculations. The NLL calculation is in fair agreement with the data while the LL one leads to a worse description.}
\label{fig_pred_d0}
\end{figure}

\subsection{Measurement at the LHC}
Both ATLAS and CMS Collaborations measured the fraction of events where there is no activity (defined as lack of jets with $p_T$ $>$ 20 GeV) between two leading jets \cite{ATLAS, CMS}. However, according to some theoretical studies (\textit{eg.} \cite{hatta}), this approach might not be sensitive to the BFKL effects. Currently, there is an effort to repeat this measurement with veto on much lower $p_T$ threshold. 

\begin{figure}[htb]
\centering
\includegraphics[width=1.0\columnwidth]{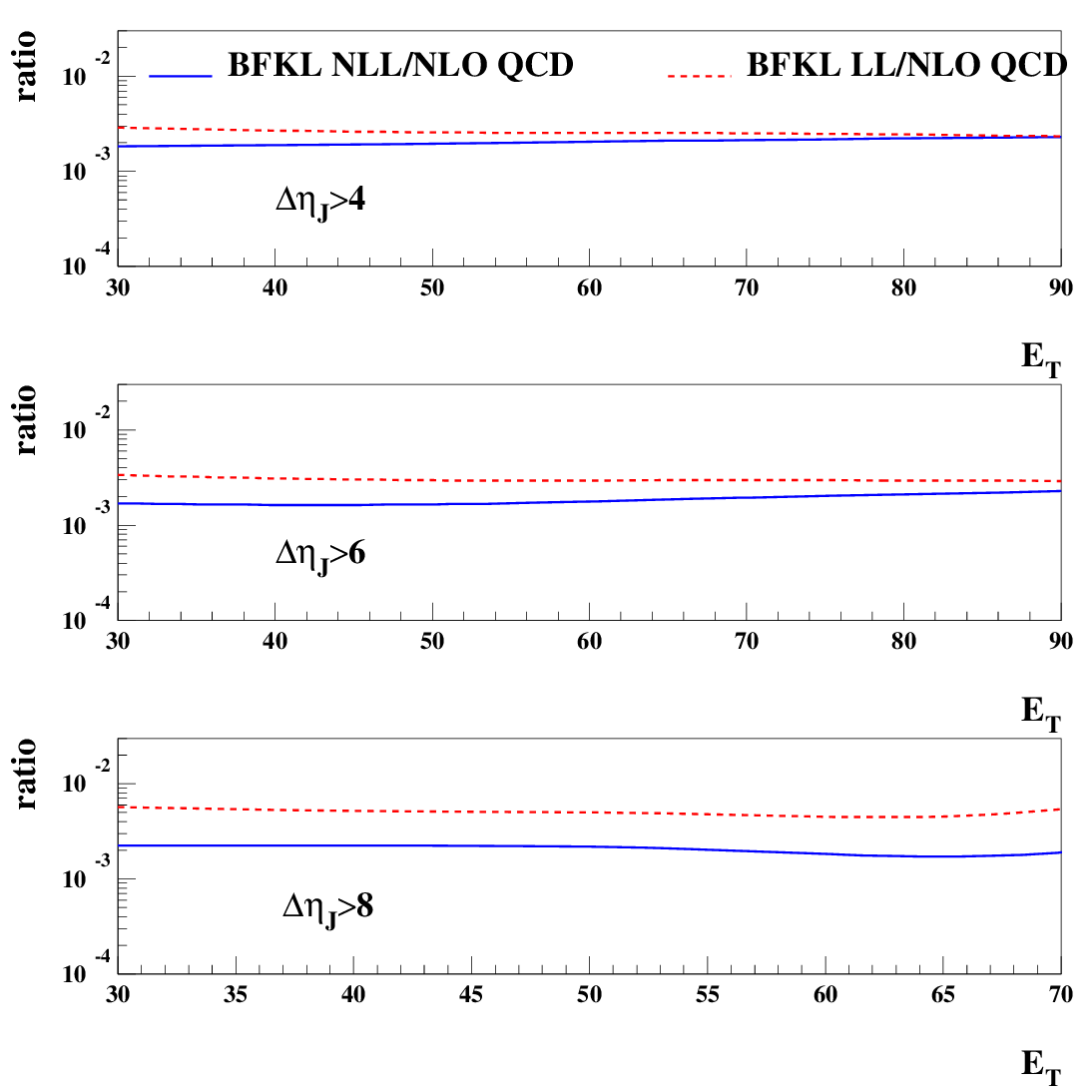}
\caption{Predictions of our model for the ratio of the jet-gap-jet to the inclusive-jet cross section at the LHC, as a function of the second-leading-jet transverse energy $E_T$.}
\label{fig_pred}
\end{figure}

The predictions of the Jet-Gap-Jet event ratio with veto on generated particles at the LHC assume the following:
\begin{itemize}
  \item the gap survival probability of 0.03,
  \item two leading jets on the opposite pseudorapidity hemispheres,
  \item the transverse energy of the second leading jets, $E_T > 20$ GeV.
\end{itemize}
The results, plotted in Figure \ref{fig_pred}, feature that the ratio is about 0.002 and does not vary a lot with $E_T$.

\section{Double Pomeron Exchange Jet-Gap-Jet Production}
\begin{figure}[!htbp]
\centering
\includegraphics[width=0.6\columnwidth]{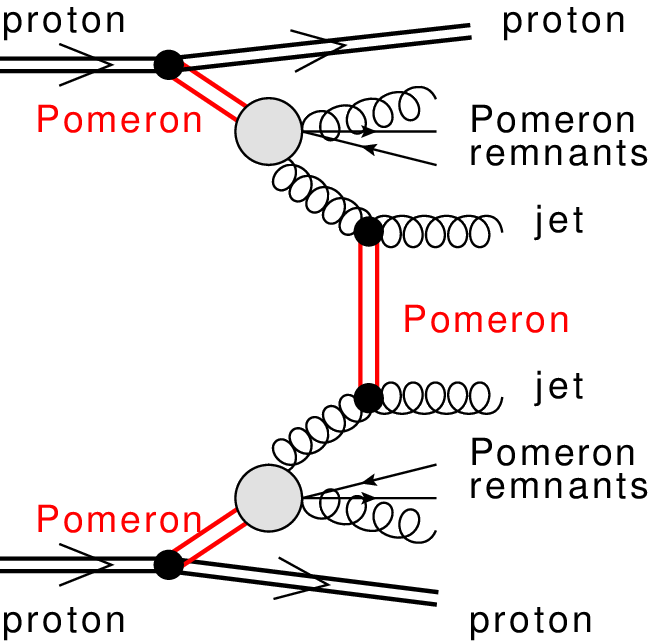}\\[0.2cm]
\includegraphics[width=0.9\columnwidth]{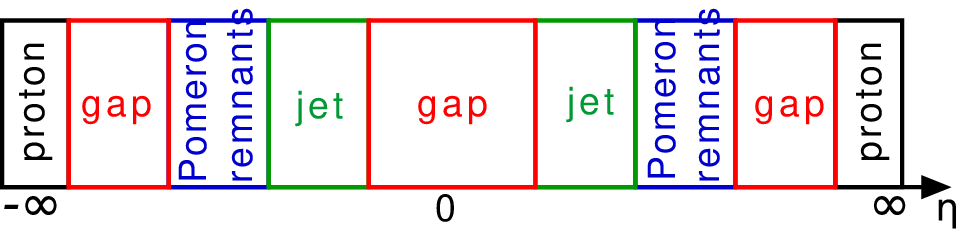}\hfill
\caption{The Double Pomeron Exchange Jet-Gap-Jet production: the interacting protons stay intact and two jets are produced. An object exchanged in $t$-channel is a colour singlet, therefore there is a gap in rapidity between two jets.}
\label{fig_diag_DPEJGJ}
\end{figure}

The discussed test of the BFKL Pomeron can also be performed in a Double Pomeron Exchange (DPE) type process (see Fig \ref{fig_diag_DPEJGJ}), hereafter called the DPE Jet-Gap-Jet. In DPE processes interacting protons stay intact, are scattered at very small angles (of the order of microradians) and stay within the accelerator beam pipe. The phenomenological formulation of the process as well as the implementation into the Monte Carlo generator is described in details in \cite{Trzebinski}.

\begin{figure}[htb]
\centering
\includegraphics[width=1.0\columnwidth]{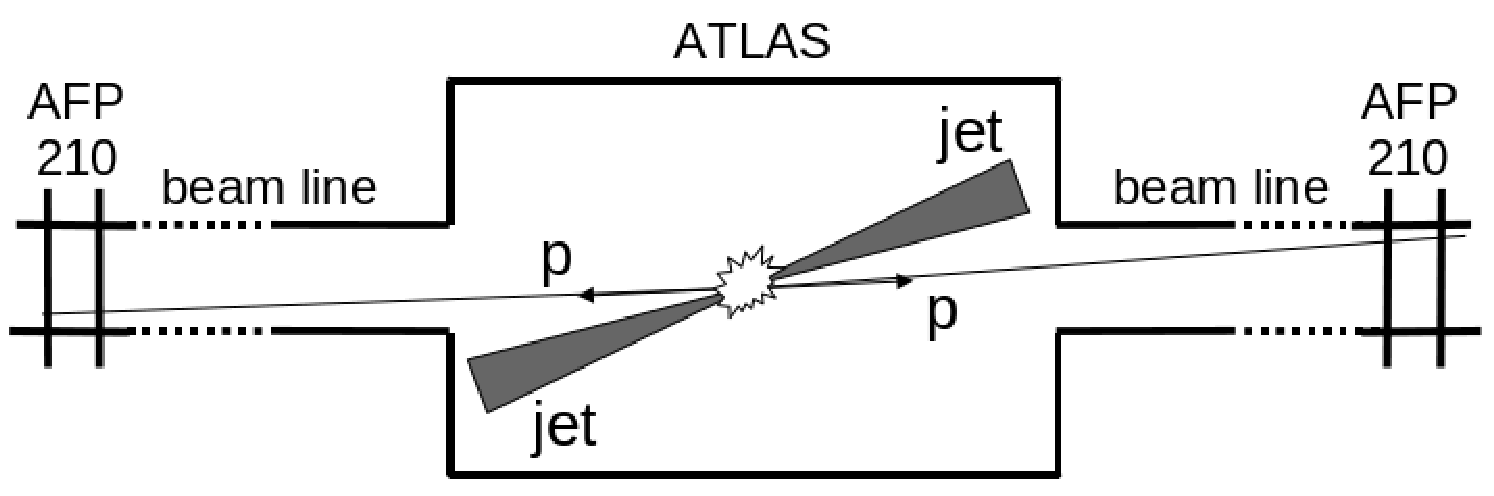}
\caption{A scheme of the measurement concept -- jets are registered in the central detectors, whereas protons in the very forward ones.}
\label{fig_scheme}
\end{figure}

To perform the DPE Jet-Gap-Jet measurement, two kinds of detectors are needed (\textit{cf.} Fig.~\ref{fig_scheme}): the central detector (for the jets detection) and the very forward detectors (for the proton tagging). The presented analysis assumes ATLAS as the central detector and the AFP (ATLAS Forward Physics) as the proton tagging devices.

\subsection{Forward Protons}
\begin{figure}[htb]
\centering
\includegraphics[width=.9\columnwidth]{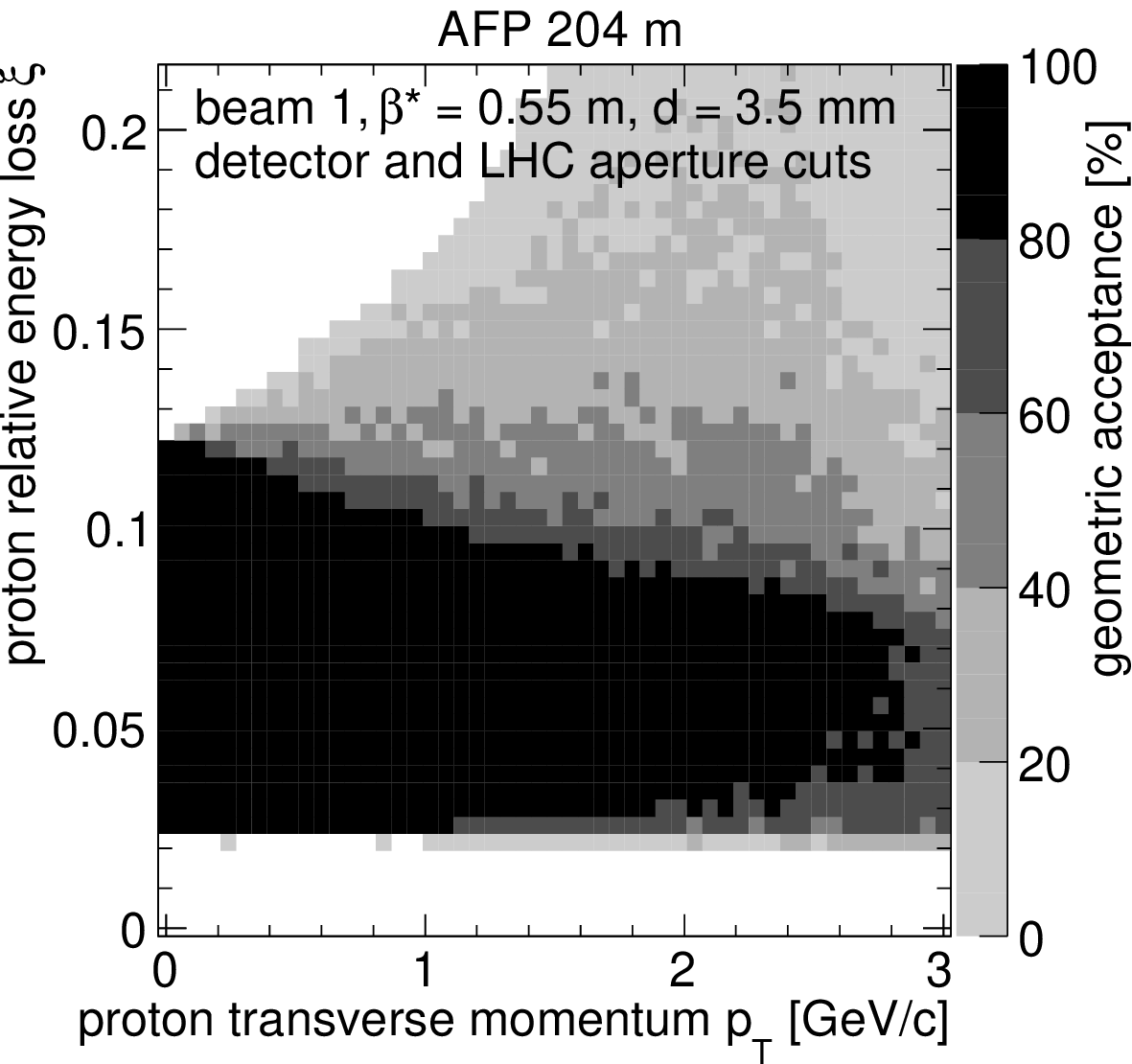}
\caption{The geometrical acceptance of the AFP detector as a function of the proton relative energy loss, $\xi$, and its transverse momentum ($p_{T}$).}
\label{fig_acceptance}
\end{figure}

Since there are several LHC magnets between the ATLAS Interaction Point and the AFP detectors, the proton trajectory depends not only on the scattering angle but also on the proton energy. Obviously, not all forward protons can be measured in the AFP detectors. Such proton can be either too close to the beam to be detected or it can hit one of the LHC elements (a collimator, the beam pipe) before it reaches the forward detector. The AFP geometric acceptance is shown in Fig. \ref{fig_acceptance}. In the calculation the following factors were taken into account: 
\begin{itemize}
  \item the beam properties at the IP, 
  \item the beam chamber geometry,
  \item the distance between the detector edge and the beam centre.
\end{itemize}
As can be observed, the region of acceptance is approximately limited by $0.012 < \xi < 0.14$ and $p_{T} <$ 4~GeV/c, where $\xi = (1 - E/E_{\mathrm{beam}}$) is the relative energy loss and $p_{T}$ is the proton transverse momentum.

Since both protons need to be tagged in the AFP stations, not all events can be recorded. The visible cross section depends on the distance between the AFP active detector edge and the beam centre. In this paper a distance of 3.5 mm is assumed, which results in a visible cross section of about 1 nb (for a leading jet heaving $p_T\ >$ 40 GeV).

\subsection{Central Diffractive Jets}
The jets produced in the DPE JGJ process will be measured in the ATLAS central detector. To fulfil the ATLAS detector trigger, the leading jet is requested to have a transverse momentum greater than 40 GeV. This rather low value is realistic for the low pile-up LHC runs required (due to the gap reconstruction) to make this measurement possible. In the presented analysis the two leading jets are required to be in the opposite pseudorapidity hemispheres and the rapidity gap is required to be symmetric around zero.

\begin{figure}[htb]
\centering
\includegraphics[width=.83\columnwidth]{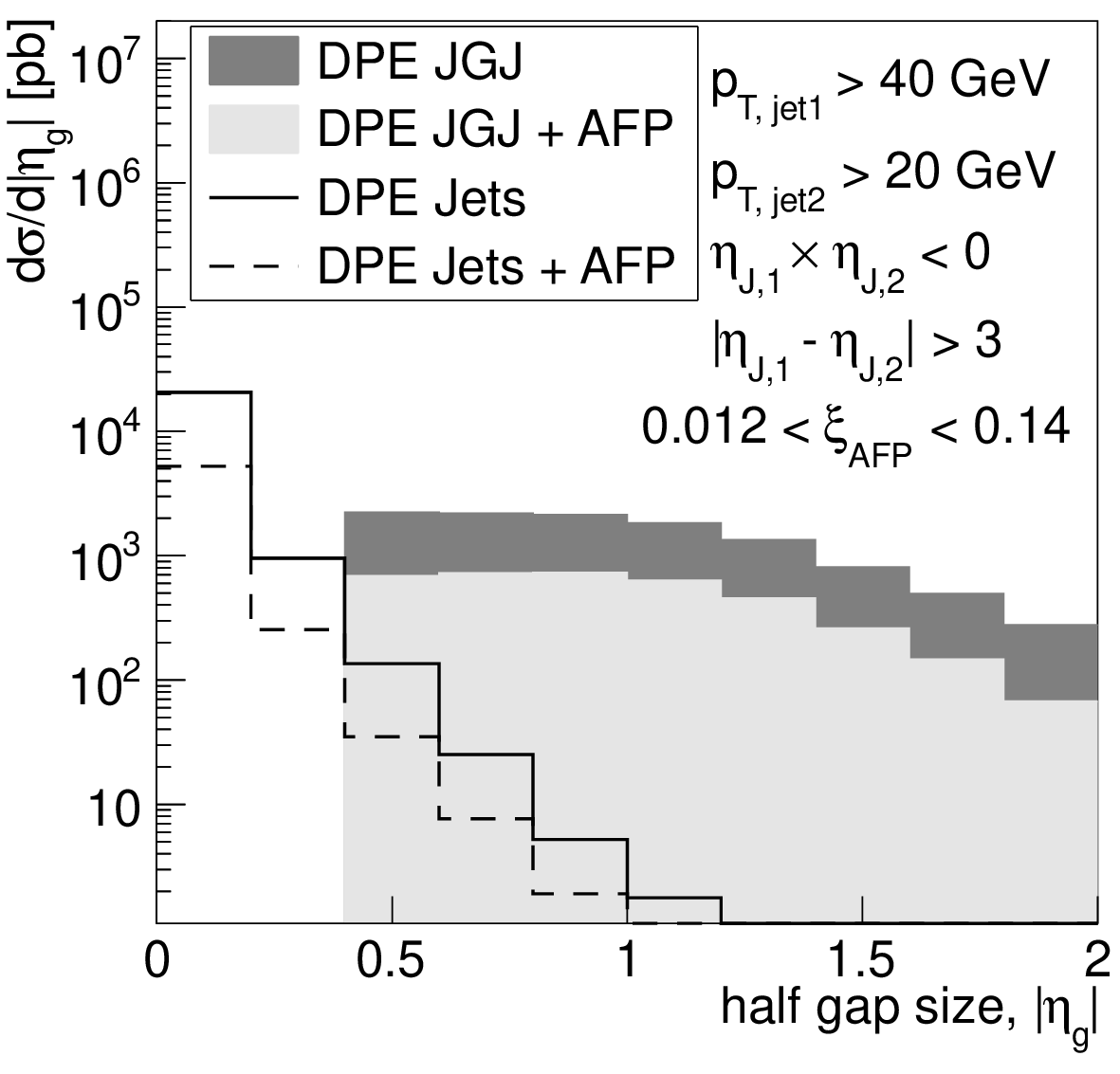}
\caption{The gap size distribution for DPE jets and DPE jet-gap-jet events with and without the AFP tag requirement.}
\label{fig_gap_dist}
\end{figure}

The main background for the DPE Jet-Gap-Jet production will be DPE inclusive jet production. In the latter process a gap between the jets can appear due to the fluctuations, but this background is significantly reduced by requiring large enough gap sizes (see Fig \ref{fig_gap_dist}). The bigger is the gap size, the larger is the DPE Jet-Gap-Jet contribution. On the other hand, the cross-section falls steeply with the increase of the gap size.

\subsection{Test of the BFKL Model at the LHC}
\begin{figure}[htb]
\centering
\includegraphics[width=.83\columnwidth]{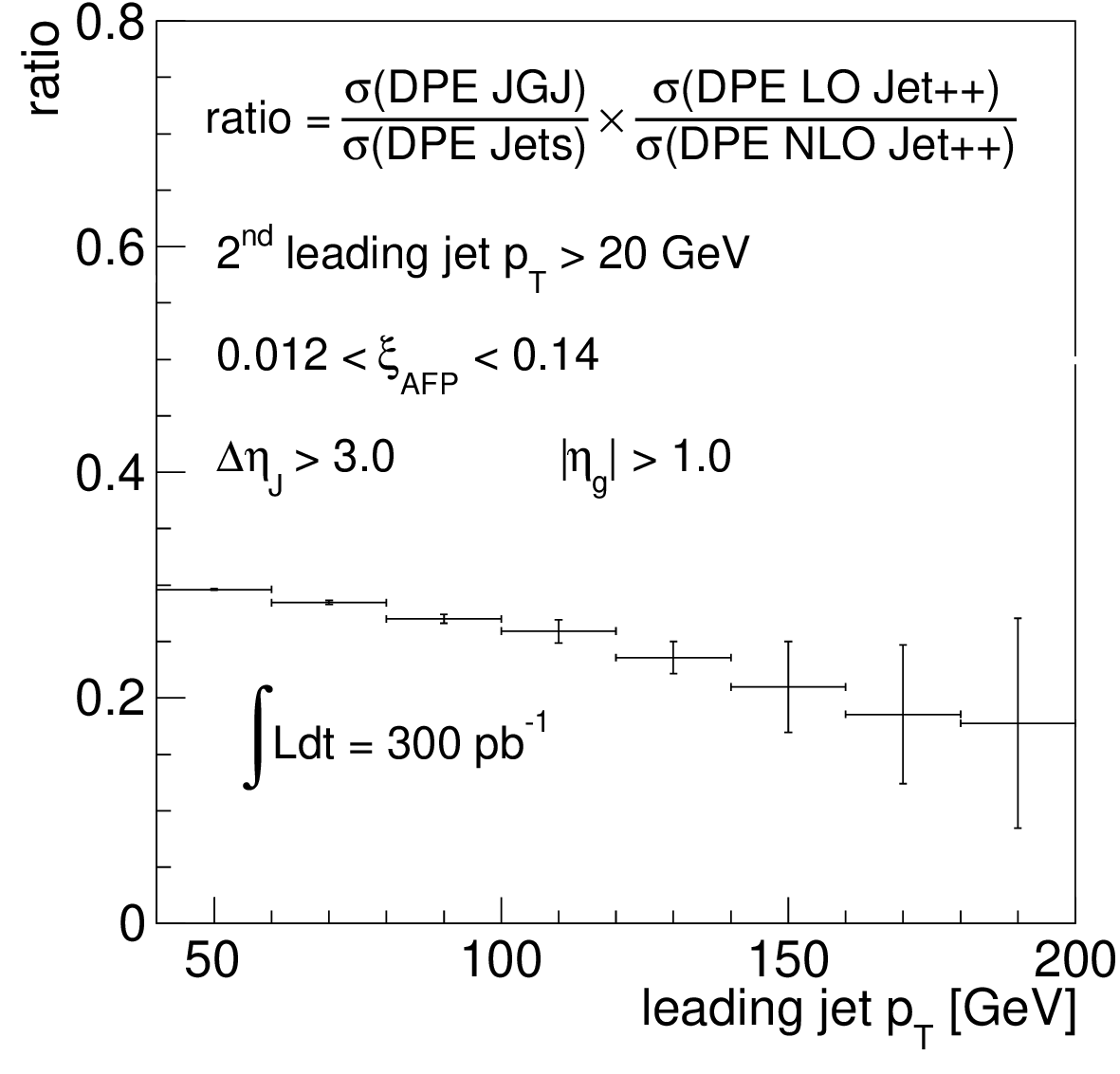}
\caption{Predictions for the DPE jet-gap-jet to DPE jet cross section ratio at the LHC, as a function of the leading jet transverse
momentum, $p_T$.}
\label{fig_dpe_pred}
\end{figure}

Similarly as in the JGJ case, one can define the DPE Jet-Gap-Jet event ratio:
$$R = \frac{\sigma(DPE\ JGJ)}{\sigma(DPE\ Jets)}.$$
This ratio is plotted in Figure \ref{fig_dpe_pred} as a function of the transverse momentum of the first-leading jet. To verify the power of this test, the statistical errors corresponding to 300 $pb^{-1}$ of integrated luminosity were plotted. 

As far as the gap fraction, $R$, is concerned there is no need to consider an additional suppression factor for DPE Jet-Gap-Jet production on top of the 0.03 of DPE inclusive jet production. Therefore, in the predictions of Fig. \ref{fig_dpe_pred}, the rapidity gap survival probabilities cancel. 

\section{Summary}
The measurement of the Jet-Gap-Jet event ratio, defined as the ratio of the Jet-Gap-Jet cross section to the inclusive di-jet cross section was performed at the Tevatron. The NLL BFKL predictions seem to be in a good agreement with the data, but due to the large statistical uncertainties the repetition of this measurement at the LHC is desirable.

At the LHC, both ATLAS and CMS measured the fraction of events where there is no activity (defined as lack of jets with the transverse momentum greater than 20 GeV) between two leading jets. However, according to \textit{eg.} \cite{hatta}, this approach might not be sensitive to the BFKL effects. The predictions for the LHC using veto on lower $p_T$ threshold are that this ratio is about 0.002 and does not vary a lot with the energy of the leading jet.

The colour singlet exchange in the $t$-channel can also be observed in the DPE processes, which provides cleaner events not polluted by proton remnants, and consequently also gives access to events with a larger rapidity difference, for which BFKL effects are more important. In addition, the fraction of Jet-Gap-Jet to inclusive di-jets events in DPE processes is larger than the corresponding fraction in non-diffractive processes, since the ratio is not penalized by the gap survival probability.

The predictions for the LHC, with ATLAS as the central detector and the AFP as the proton tagging devices, are that there will be enough statistics in 300 $pb^{-1}$ of data to collect a significant sample of DPE Jet-Gap-Jet events. This would provide an additional test of the BFKL Pomeron.

\section{Acknowledgements}
This work was supported by the Polish National Science Centre grant number UMO-2012/05/N/ST2/02697.

\end{document}